\documentstyle[11pt,newpasp,twoside]{article}
\def\edcomment#1{\iffalse\marginpar{\raggedright\sl#1\/}\else\relax\fi}
\reversemarginpar

\begin{document}
\title{Constraining the Heliosphere:  The Need for High-Resolution Observations of Nearby Interstellar Matter} 
\author{Priscilla C. Frisch}

\begin{abstract}
High-resolution ultraviolet observations of nearby bright and faint
stars are required to evaluate changes in the past and future galactic
environments of the Sun, and the possibly impact of these changes on
the interplanetary environment at 1 AU (around the Earth).  The
boundary conditions of the heliosphere and interplanetary environments
are constrained by the characteristics of the surrounding interstellar
material (ISM), which changes on timescales of 10$^3$--10$^5$ years.
An increase in the density of the interstellar cloud surrounding
the solar system to 10 cm$^{-3}$ decreases the
heliosphere radius by about an order of magnitude.  UV observations of
nearby stars at high spectral resolution ($>$300,000) and high signal-to-noise are required to
evaluate future modifications to heliosphere properties by the ISM.
\end{abstract}

\section{Introduction}

High-resolution ultraviolet observations of nearby bright and faint
stars are required to evaluate changes in the past and future galactic
environments of the Sun, and the possibly impact of these changes on
the interplanetary environment at 1 AU (around the Earth) and
in the outer solar system.

\section{The Heliosphere, Interstellar, and Interplanetary Matter}

The interplanetary environment around the Earth and the boundary
conditions of the heliosphere depend on the physical properties of the
interstellar cloud surrounding the solar system (known as the ``Local
Interstellar Cloud'', or LIC).  These boundary conditions vary with
the space motions of the Sun 
\footnote{The LSR velocity used here is the Hipparcos value,
corresponding to a solar motion towards the galactic coordinates
l=27.7\deg, b=32.4\deg\ at the velocity $V$=13.4 km s$^{\rm -1}$
(Dehnen \& Binney 1998). }  ($\sim$13.5 km s$^{\rm -1}$) and
interstellar clouds (0--30 km s$^{\rm -1}$) through the Local Standard
of Rest (LSR).  The primary visitors from beyond the heliopause are
neutral atoms, large dust grains (radii$>$0.20 $\mu$m, Frisch et al. 1999), and cosmic
rays with energies $>$1 GeV.  Presently $\sim$98\% of the diffuse
material in the solar system is the neutral component of interstellar
gas which flows relatively freely through the heliopause region.  The
solar wind plasma and interstellar neutral densities are equal at
$\sim$6 AU because of the 1/$R^2$ falloff in solar wind density.  The
solar wind shields Earth from the low pressure and low density
interstellar cloud around and within the solar system ($P_{\rm
thermal} \sim$2500 cm$^{-3}$ K).  Variations in the properties of the
cloud surrounding the solar system yield variations in the heliosphere
configuration and the amount of ISM in the inner solar system.

The heliopause location and heliosphere configuration depend on the
relative ram pressures of the solar wind and ISM (Holzer 1989).  The
current boundary conditions of the heliosphere are a relative
Sun-cloud velocity of $\sim$26 km s$^{-1}$, $n$(HI)$\sim$0.24
cm$^{-3}$, $n$(e)$\sim$0.13 cm$^{-3}$, T$\sim$7,000 K, $X$(H)=0.31 and
$X$(He)=0.48, based on radiative transfer models of ISM within 3 pc,
and pickup ion and ISM absorption data (Slavin \& Frisch 2002).  Today
the heliosphere radius is large ($\sim$150 AU) and varies with the solar cycle.

Increasing the ISM neutral density to 10 atoms cm$^{-3}$ has been
shown to contract the heliopause radius by an order of magnitude
and the heliopause becomes Rayleigh-Taylor unstable from solar wind mass-loading (Zank
\& Frisch 1999).  This change in the heliosphere size would modify the
1 AU interplanetary environment of the Earth.  
The flux of interstellar
neutrals at the Earth's orbit increases to $\sim$2 cm$^{-3}$, and the
heliopause radius contracts to $R \sim$10--14 AU, and
outer planets are exposed to the raw ISM, dramatically
altering the interplanetary environment of the outer solar system.

\section{Variations in Properties of ISM Surrounding the Solar System}

The Sun is embedded in a cluster of interstellar cloudlets flowing
outwards from the region of the Loop I superbubble.  High resolution
ground observations of CaII (resolution 0.3--1.0 km s$^{\rm -1}$) show
that the ISM flow past the solar system contains at least six distinct
velocity component groups suggestive of separate clouds.  About 96
components are observed towards 60 nearby stars.  Several of the
nearest stars show $\sim$1 component per 1.4--1.6 pc.  CaII component
velocities towards nearby stars exhibit random velocities (rms
dispersion $\sigma \sim$5 km s$^{\rm -1}$) distributed about a bulk
flow velocity with upstream direction towards Loop I.  In the LSR, the
bulk flow velocity corresponds to a motion of --17.0 km s$^{\rm -1}$
with an upstream direction of l=2.3\deg, b=--5.2\deg\ (Frisch et al. 2002, F02).

Cloud velocity is a proxy for cloud structure.  Typical 
relative Sun-ISM velocities of
0$\rightarrow$40 km s$^{\rm -1}$ (or more) are to be expected.  An
encounter with a `new' cloud is thus possible every $\sim$25,000
years, and the potential for variations in the Galactic
environment of the Sun exists for even shorter timescales.

Variable boundary conditions yield variable heliosphere properties.
One hundred thousand years ago the Sun was immersed in the interior of the
Local Bubble (T$\sim 10^6$ K, n$_{\rm p} \sim$0.005 cm$^{-3}$),
and the heliosphere radius was
similar to today but with no interstellar neutrals inside of the solar
system (e.g. Mueller et al. 2002).  The Sun must have emerged
from the Local Bubble interior and entered the cloud now surrounding
the Sun 2,000--10$^5$ years ago (Frisch 1994).  As the Sun continues to
travel through the outflow of cloudlets from the Loop I region, new
cloudlets will be encountered.  

The nearest star $\alpha$ Cen (1.3 pc) is located near the LIC LSR upstream direction (F02), however the
velocity of ISM towards $\alpha$ Cen differs by several km s$^{\rm
-1}$ from the projected LIC velocity suggesting the Sun may encounter
the $\alpha$ Cen cloudlet within 10$^4$ years.  Another cloudlet with
d$<$5 pc and a heliocentric velocity of $\sim$30 km s$^{\rm -1}$
is located in the solar apex direction.  If this cloud has 0 km s$^{\rm
-1}$ tangential velocity, the Sun will encounter this cloud within
$\sim$1--2 x 10$^5$ years (F02).

\section{Advantage of High-Resolution UV data}

Evaluating possible changes in the boundary conditions of the
heliosphere requires UV observations of the individual cloudlets near
the Sun, since the required resonance transitions of dominate ions are
not accessible from the ground.  The tag which identifies individual
cloudlets is component velocity, and high resolution optical CaII data
show the component structure of interstellar clouds is
undersampled in the UV at resolutions of 3 km s$^{\rm -1}$.  Welty et
al. (1996) have shown a crowding of optical components in velocity
space, such that the number of detected components increases exponentially with
improved instrumental resolution.  As a result, observations at the resolution of HST GHRS
and STIS appear to miss over 60\% of the absorption components.
High resolution and high signal-to-noise observations in
low column density sightlines ($<$10$^{18}$ cm$^{-2}$) towards nearby
stars provide the best opportunity to resolve individual cloudlets free from 
the well-known but difficult-to-specify uncertainties resulting from blended components.  

Optical observations of CaII towards the star $\alpha$ Oph (14 pc) provide an
example of the gains expected from improved UV resolution.  For resolution $\sim 10^6$, UHRS data show four absorption
components, compared to three components found at the lower resolution
of 250,000 (e.g. Crawford \& Dunkin 1995, Welty et al. 1996, Crawford 2001).  
Doppler b-value varies from 1.06 to 4.04 for the four components (or T=2700 K to
39000 K, if no turbulence, Crawford 2001).  
The weak CaII feature towards $\alpha$ Oph  at --31.8 km s$^{-1}$ is formed in a cloud within 5 pc of the Sun (F02),
and the b-value is alternately reported as 1.5 km s$^{-1}$ and
4.0 km s$^{-1}$ in the analysis of UHRS data.  This difference emphasizes
that high resolution data must also be high signal-to-noise data to remove
uncertainties in interpretating observations of very low column density
cloudlets.  

Separating turbulence from thermal
broadening requires UV observations with resolutions $\sim$1 km
s$^{-1}$, and understanding the ionization and densities of the
individual components requires resolving the component structure for a
range of ions (e.g. FeII, MgII, MgI).  Resolving out possible cold
(100 K) cloudlets embedded in the flow of ISM past the Sun requires
resolutions on the order of 10$^6$.  Such high resolution will permit
determining the thermal width of MgI lines in cool neutral (100 K, b=0.26 km s$^{-1}$)
versus warm partially ionized (T=7000 K, b=2.2 km s$^{-1}$) cloudlets,
so that the physics of MgI in the local ISM (i.e. radiative versus dielectronic recombination)
can be resolved.

Factor of two or greater variations in FeII abundances are seen in cloudlets
observed towards nearby stars, showing that small-scale
structure and pressure variations remain to be discovered in the 
immediate galactic environment of the Sun with high-resolution UV data.

\section{Conclusions}

The morphology of nearby ISM ($<$ 30 pc), and interstellar pressure
variations affecting the heliosphere and interplanetary medium
surrounding the Earth, {\it can only be found} from high-resolution
($\lambda$/$\delta \lambda$$>$300,000) observations of interstellar
absorption lines in the ultraviolet wavelength band (912 --- 3000
Ang).  The relative simplicity of sightlines to nearby stars,
combined with high-resolution high signal-to-noise UV data,
provides the best opportunity for obtaining accurate abundances
and cloud properties for interstellar clouds.


\begin{references}
\reference {Crawford}, I.~A. \& {Dunkin}, S.~K. 1995, \mnras, 273, 219 
\reference {Crawford}, I.~A. 2001, \mnras, 327, 841 
\reference {Dehnen}, W. \& Binney, J.~J. 1998, \mnras, 298, 387 
\reference {Frisch}, P.~C. 1994, Science, 265, 1423 
\reference {Frisch}, P.~C., {Dorschner}, J.~M., {Geiss}, J., {Greenberg}, J.~M., {Gr\"un},
  E., {Landgraf}, M., {Hoppe}, P., {Jones}, A.~P., {Kr{\"{a}}tschmer}, W.,
  {Linde}, T.~J., {Morfill}, G.~E., {Reach}, W., {Slavin}, J.~D., {Svestka},
  J., {Witt}, A.~N., \& {Zank}, G.~P. 1999, \apj, 525, 492
\reference Frisch, P.~C., Grodnicki, L., \& Welty, D.~E. 2002, \apj, August issue (F02)
\reference {Holzer}, T.~E. 1989, \araa, 27, 199 
\reference {Mueller}, H.~R., {Frisch}, P.~C., \& {Zank}, G.~P. 2002, in preparation
\reference {Slavin}, J.~D. \& {Frisch}, P.~C. 2002, \apj, 565, 364
\reference {Welty}, D.~E., {Morton}, D.~C., \& {Hobbs}, L.~M. 1996, \apjs, 106, 533 
\reference {Zank}, G.~P. \& {Frisch}, P.~C. 1999, \apj, 518, 965 
\end{references}
\end{document}